\documentclass{elsarticle}
\usepackage{times}
\usepackage{amsfonts}
\usepackage{mathptmx}
\usepackage{bm}
\newcommand{\beq}{\begin{equation}}
\newcommand{\eeq}{\end{equation}}
\newcommand{\beqa}{\begin{eqnarray}}
\newcommand{\eeqa}{\end{eqnarray}}
\newcommand{\nn}{\nonumber \\}

\def \e {\mathrm{e}}

\def \el {\mathrm{el}}

\def \imb {\mathrm{imb}}

\def \la {\langle}
\def \ra {\rangle}
\def \s {\sigma}
\def \t {\tau}

\def \P {{\mathcal P}}
\def \R {{\mathbb R}}
\def \Z {{\mathbb Z}}
\def \ch {\mathrm{ch}}

\def \z {\zeta}
    \def \D {\Delta}

\def \Im {\mathrm{Im} \, }

\def \mod {\ \mathrm{mod} \ }
\def \H {{\mathcal H}}
\def \uu {{\widehat{u(1)}}}

\begin{document}
\begin{frontmatter}
\title{Thermoelectric properties of Coulomb-blockaded fractional quantum Hall islands}
\author{Lachezar S. Georgiev}
\ead{lgeorg@inrne.bas.bg}
\address{Institute for Nuclear Research and Nuclear Energy, Bulgarian Academy of Sciences,
	72 Tsarigradsko Chaussee, 1784 Sofia, Bulgaria}

\begin{abstract}
We show that it is possible and rather efficient to compute at non-zero temperature  the thermoelectric characteristics of 
Coulomb blockaded fractional quantum Hall islands, formed by two quantum point contacts inside of a Fabry--P\'erot interferometer,
using the conformal field theory partition functions for the chiral edge excitations. The oscillations of the thermopower 
with the variation of the gate voltage as well as the corresponding figure-of-merit and power factors,
provide finer spectroscopic tools which are sensitive to the neutral multiplicities in the partition functions and could be used to 
distinguish experimentally between different universality classes sharing the same electric properties.
We also propose a procedure for measuring  the ratio $r=v_n/v_c$ of the Fermi velocities of the neutral and charged 
edge modes for filling factor $\nu_H=5/2$ from the power-factor data in the low-temperature limit.
\end{abstract}
\end{frontmatter}
\section{Introduction}
Investigating the thermoelectric properties of strongly correlated two-dimensional electron systems is expected to reveal 
important information about the structure of  the neutral excitations \cite{viola-stern} and other specific characteristics of their 
universality classes. Distinguishing between different candidate states describing fractional quantum Hall (FQH) universality classes 
is interesting because some of them are expected to have particle-like excitations obeying non-Abelian exchange (or braid) 
statistics \cite{blok-wen,mr,rr}. Besides the fundamental importance of non-Abelian quasiparticles as new types of particles with exotic 
statistics, which could only exist in two dimensions, they are also believed to play a crucial role in the field of topological 
quantum computation where the strange but very robust braid statistics of the quasiparticles in combination with the 
topology of the quantum registers could efficiently protect quantum information against noise and 
decoherence \cite{sarma-RMP,stern-review}.

In an attempt to distinguish between the different candidates for the $\nu=5/2$ FQH state, people have investigated the 
Coulomb-blockade (CB) conductance patterns of FQH  islands in states from different universality classes, including at 
non-zero temperature \cite{stern-CB-RR,stern-CB-RR-PRB,cappelli-viola,thermal}.
Unfortunately  the CB data appears to be insufficient at low temperature \cite{nayak-doppel-CB} for distinguishing different states 
because FQH states from different universality classes have been shown to have identical  CB conductance peak patterns
at zero temperature.

Recently, an emerging possibility to detect non-Abelian statistics by measuring the thermoelectric properties 
of  different FQH states in the CB regime of a Fabry-P\'erot interferometer \cite{viola-stern} has attracted some attention.
The thermoelectric conductance of candidate FQH states  at filling factors $\nu_H=2/3$ and $\nu_H=5/2$ in a CB island 
have been computed  \cite{viola-stern} from the conformal field theory (CFT) 
data of the underlying effective field theories for the edge excitations. It has been demonstrated that  thermoelectric conductance
of the quantum dot formed inside of the Fabry-P\'erot interferometer might be sensitive to the neutral degrees of freedom of the FQH 
states expressed in eventually measurable asymmetries for even number of localized quasiparticles in the bulk. However,
it appeared that the computation of the thermoelectric conductance for $\nu_H=5/2$ strongly depends on the ratio 
$r=v_n/v_c$ of the Fermi velocities of the neutral and charged edge modes, which has to be considered as a free parameter. 
In Ref.~\cite{viola-stern} the value $r\approx 1/6$ has been chosen with the argument that it is consistent with previous 
numerical and experimental work.

Another thermoelectric quantity, the thermopower, known also as the Seebeck coefficient, 
has been previously computed for metallic quantum dots \cite{matveev-LNP,beenakker-staring-thermopower} indicating to be a better 
spectrometric tool than the transport coefficients alone, while showing the same periodicity as the CB conductance peaks. 
So far, the computations of thermopower for CB islands in quantum Hall states have been limited to the case of integer $\nu_H$
where it is similar to that of the metallic islands. Recently, the thermopower for the $\nu_H=1/m$ Laughlin FQH states has 
been computed \cite{LT10} showing that it is similar to the integer quantum Hall states, except that the oscillation period
 in the dimensionless Aharonov-Bohm flux (related to the gate voltage) is extended from $1$ to $m$.
 
The chiral edge excitations determining the topological order of the FQH universality classes have been successfully described by 
CFTs \cite{wen,fro-ker,ctz,read-CFT}.
In this paper we will show how to use the CFT partition function for a general chiral FQH state, as a thermodynamic potential   
for the experimental setup of Refs.~\cite{thermal,viola-stern}, in order to calculate the thermopower for a CB island, 
or a quantum dot (QD), at non-zero temperature. Measuring the power factor computed from the thermopower could experimentally
help us to estimate the ratio $r=v_n/v_c$ of the Fermi velocities of the neutral and charged edge modes and eventually to
distinguish between the different $\nu_H=5/2$  states.

The rest of this paper is organized as follows: in Sect.~\ref{sec:tunneling} we explain how the thermopower of a Coulomb 
blockaded fractional quantum Hall island can be expressed in terms of the Grand-canonical averages of the 
edge states' Hamiltonian and particle number operators. In Sect.~\ref{sec:disk} we review the structure of the Grand-canonical 
partition functions for general FQH states on a disk and discuss how they are modified in presence of Aharonov--Bohm flux.
In Sect.~\ref{sec:5-2} we consider as an example the proposed paired states, the Pfaffian state, the Halperin 331 state, 
the $SU(2)_2$ and the anti-Pfaffian state,  which are candidates to describe the universality class of the fractional quantum Hall 
state at filling factor $\nu_H=5/2$. We calculate numerically and plot the thermopower, the electric and thermal conductances 
and the power factors for these states with odd or even number of bulk quasiparticles. We finish by a discussion of the open 
problems related to the description of FQH states with counter-propagating modes and give some additional information in 
three appendices.
\section{Thermopower as average tunneling energy of FQH edge excitations}
\label{sec:tunneling}
\begin{figure}[htb]
\centering
\includegraphics[bb=40 300 560 535,clip,width=\textwidth]{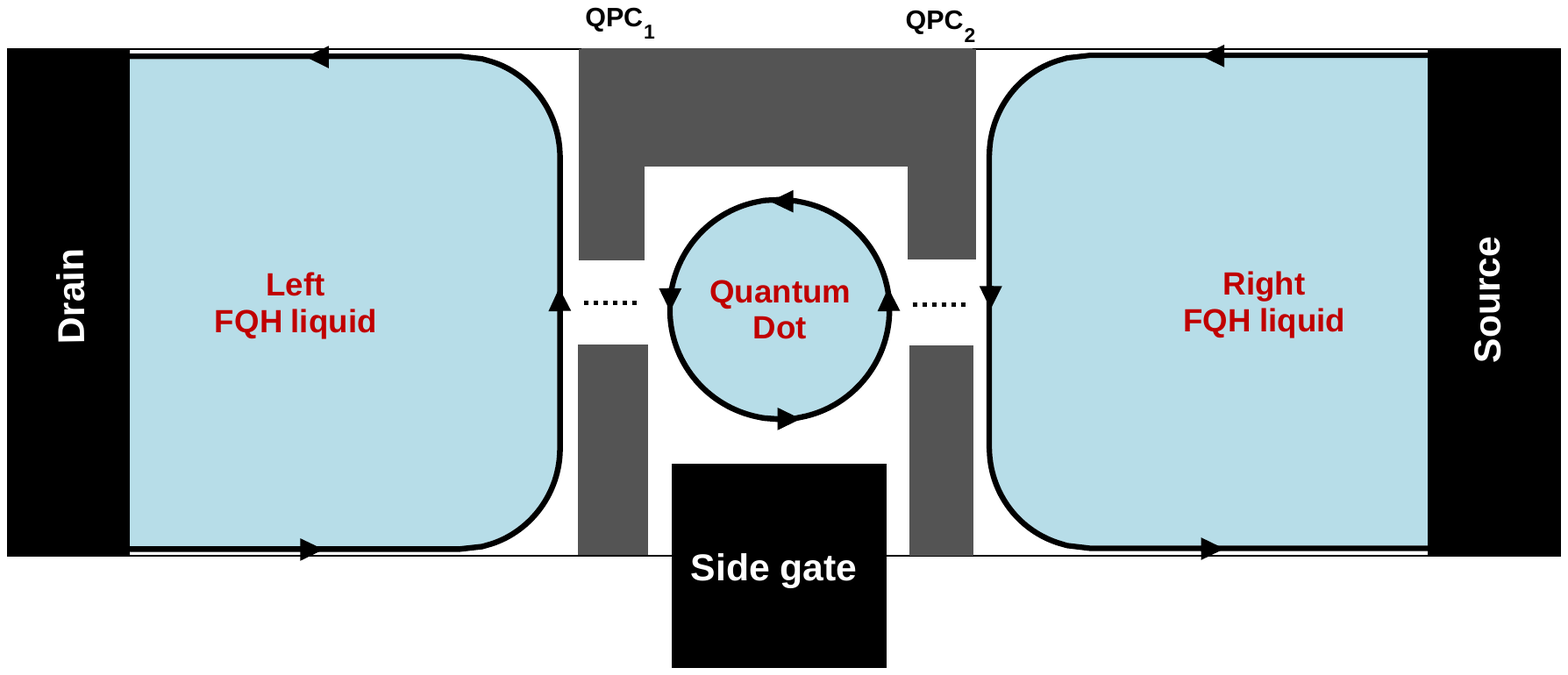}
\caption{Single-electron transistor realized by two quantum-point contacts (QPC$_1$ and QPC$_2$) inside of
 a FQH bar. The arrows show the direction of the propagation of the edge modes. 
Only electrons can tunnel between the left and right FQH liquids and the QD under 
appropriate conditions and the dots mark the electron tunneling paths. \label{fig:SET}}
\end{figure}

The  thermopower is defined \cite{matveev-LNP} as the potential difference $V$ between the 
left and right leads of the single-electron transistor (SET), formed by the CB island (or QD) and the 
Drain, Source and Side gate as shown in Fig.~\ref{fig:SET},
when their temperature differs by $\Delta T=T_R -T_L\ll T_L$, under the condition that the 
electric current $I$ is 0. It is usually computed  \cite{matveev-LNP}  as the ratio 
\beq \label{S-G}
S=\frac{G_T}{G}
\eeq
 of the thermal and 
electric conductances, $G_T$ and $G$ respectively, however, it can be alternatively expressed in terms of the average 
energy $\la \varepsilon \ra$ of the 
electrons tunneling through the Coulomb-blockaded quantum dot \cite{matveev-LNP}
\beq \label{S}
S \equiv  \left. -\lim_{\Delta T \to 0} \frac{V}{\Delta T}\right|_{I=0}=-\frac{\la \varepsilon \ra}{eT},
\eeq
where $e$ is the electron charge and $T$ is the temperature of the CB island.
The thermopower is measured in units $V . K^{-1}$ or, as obvious from the right-hand side of Eq.~(\ref{S}),
in  units $J/(A.s.K)$ in the SI system. Because the electric conductance $G$ is measured in units $e^2/h$
and the thermal conductance $G_T$ is measured in $ek_B/h$ it follows from Eq.~(\ref{S-G}) that thermopower 
can be also measured in units $k_B/e$. 
The alternative approach based on the right-hand side of Eq.~(\ref{S}) is more suitable for disconnected systems, 
such as the SET shown in 
Fig.~\ref{fig:SET}, because the 
conductances $G_T$ and $G$ are both zero in large intervals called  CB valleys \cite{matveev-LNP}, so it is not 
appropriate to put $G$ in the denominator of (\ref{S}), while the voltage $V$ is non-zero and can be measured 
experimentally \cite{staring-exp}.
The knowledge  of the thermopower and the conductances allow us to compute  another important thermoelectric 
characteristics \cite{zt}--the thermoelectric figure-of-merit 
\beq \label{ZT}
 ZT= \frac{S^2 G T}{G_T}
\eeq 
for the CB island as a quantum dot and the corresponding power factor $\P_T$,  
which is defined in terms of the electric power $P$ generated by $\Delta T$ as
\beq\label{P_T}
P=V^2 / R = \P_T (\Delta T)^2, \quad \P_T= S^2 G,
\eeq
where $R=1/G$ is the electric resistance of the CB island.
We emphasize here that Eq.~(\ref{S-G}) is relevant only when the electric conductance
$G$ is nonzero, while the standard formulas for the figure-of-merit (\ref{ZT})  and the power-factor (\ref{P_T}) are 
still expressible in terms of the thermopower $S$ even when the ratio (\ref{S-G}) is experimentally indeterminate.
That is why thermopower carries more information about the strongly interacting electron system than the electric 
and thermal conductances together.

The power factor $\P_T$ seems to be measurable directly by applying an AC voltage of frequency $f_0/2$, to 
the side gate,  while measuring the thermoelectric current at frequency $f_0$ \cite{gurman-2-3}.
It is worth stressing that the sharp zeros  of the power factor $\P_T$, at very low temperatures,  
mark precisely the positions of the maximum of the CB conductance 
peaks  and could be used to determine experimentally the ratio $r=v_n/v_c$ of the Fermi velocities $v_n$ and $v_c$ for the 
neutral and charged modes respectively, see Eq.~(\ref{r}) below. On the other hand the ratio of the two maxima 
of the power factor $\P_T$ around each CB peak, just like the ratio of the two extrema of $G_T$ in \cite{viola-stern},
appears to be rather sensitive to the  presence of neutral degeneracies in the edge modes due to finite-temperature 
asymmetries in the conductance peaks \cite{thermal}. Moreover, it has been shown that $\P_T$ and $S$
 could be significantly enhanced in the single-electron-transistor setup due to the Coulomb blockade \cite{enhanced}.
Therefore, $\P_T$ and $S$  could  eventually be used to distinguish between different FQH universality classes having the 
same CB peak pattern  \cite{nayak-doppel-CB}.

In the rest of this section we will describe how to use the CFT for the edge excitations of a disk FQH sample
to compute thermopower and power factors of the corresponding CB islands, which is a central result in this paper. 
To this end, we first identify the average electron tunneling energy in Eq.~(\ref{S}) as the difference between the total energy of the QD 
with $N+1$  and $N$ electrons, which have the same bulk but different edge contributions, then we calculate the edge QD energy and 
edge electron number, in presence of Aharonov--Bohm flux or gate voltage, as the Grand-canonical averages of the twisted 
CFT Hamiltonian and Luttinger liquid particle number operator, respectively and finally we express these thermal averages in terms of 
the Grand canonical disk partition function derived within the framework of the CFT for the edge states.

For large CB islands, the total energy of $E_{\mathrm{QD}}$ of the QD with $N$ electrons is defined
within the \textit{Constant Interaction model} \cite{kouwenhoven} 
as\footnote{following Ref.~\cite{viola-stern} we disregard the electrostatic energy $e^2N^2/2C$ which is subleading 
for the large CB islands that are of experimental interest}
\beq \label{E_N}
E^{\beta,\mu_{N}}_{\mathrm{QD}}(\phi) =\sum_{i=1}^{N_0} E_i (B)+ \la H_{\mathrm{CFT}}(\phi)\ra_{\beta,\mu_{N}} ,
\eeq
where $N_0$ is the number of electrons in the bulk of the QD and $N-N_0=N_\el$ is the number of electrons on the edge,
 $E_i(B)$, $i=1, \ldots , N_0$, are the energies of the occupied single-electron states in the bulk of the QD.
Since we intend to use the CFT partition function of the CB island as a Grand potential 
the average $\la \cdots \ra_{\beta,\mu}$ is taken within the Grand canonical ensemble for the FQH edge at inverse 
temperature $\beta=(k_B T)^{-1}$ and chemical potential $\mu$ and $\la H_{\mathrm{CFT}} \ra_{\beta,\mu}$ is the 
Grand canonical average of the Hamiltonian on the edge. 
At low temperature the number of electrons on the QD is 
quantized to be integer and the chemical potential of the QD with $N$ electrons is 
denoted by $\mu_N$.

We will be interested in the sequential tunneling regime \cite{matveev-LNP,thermal}, 
when the electrons in the SET tunnel through the QD one by one as the gate voltage is varied,
which is the dominating mechanism at low temperature for small conductances between the CB island and the ``leads'',
like in \cite{matveev-LNP,thermal}. The leads are assumed to be large FQH liquids with energy spacing much smaller than the energy 
spacing $\Delta\varepsilon$ of the CB island.  
In this case, within the linear response approximation for low temperature differences $\Delta T$,  the average energy of the 
electrons tunneling through the CB island could be computed as  the difference between the  average total energy of the 
QD with $N+1$ electrons and that for $N$ electrons in presence of AB flux $\phi$ (or, equivalently, normalized gate voltage)
\beq\label{eps}
\la \varepsilon \ra^{\phi}_{\beta,\mu_N} =
\frac{\la H_{\mathrm{CFT}}(\phi)\ra_{\beta,\mu_{N+1}} - \la H_{\mathrm{CFT}}(\phi)\ra_{\beta,\mu_N}}
{\la N_\el(\phi)\ra_{\beta,\mu_{N+1}} - \la N_\el(\phi)\ra_{\beta,\mu_N}}.
\eeq
The variation of the side-gate voltage $V_g$ induces (continuously varying) ``external charge''
$e N_g=C_g V_g$ on the edge \cite{kouwenhoven,staring-CB},  which is equivalent to the AB flux-induced variation of the particle number 
$N_\phi=\nu_H \phi$, so that we can use instead the AB flux $\phi$ determined from
 $C_gV_g/e\equiv \nu_H\phi$ with $\phi=(e/h)\left( BA-B_0A_0\right)$,
where $A_0$ is the area of the CB island and $B_0$ is the magnetic field at $V_g=0$. 
Using the AB flux $\phi$ instead of the gate voltage is convenient because $\phi$  
can be interpreted mathematically as a continuous twisting of the $\uu$ charge of the underlying chiral 
algebra  \cite{NPB-PF_k,CFT-book}, which is technically similar to the rational (orbifold) twisting of the $\uu$ 
current  \cite{kt}.
All averages entering Eq.~(\ref{eps}) could be identified with some derivatives of the thermodynamical Grand potential. 
The Grand potential $\Omega(\beta,\mu)=-\beta^{-1} \ln Z(\beta,\mu)$, for the FQH edge states, is defined as usual \cite{kubo}  
in terms of the Grand canonical partition function
\beq\label{Z0}
Z(\beta,\mu)= \mathrm{tr}_{\H} \e^{-\beta (H_{\mathrm{CFT}} -\mu N_\el)},
\eeq
where $H_{\mathrm{CFT}}=\Delta\varepsilon (L_0-c/24)$ is the Hamiltonian for the edge states expressed in terms of the 
zero mode $L_0$ of
the Virassoro  stress tensor \cite{CFT-book} (with central charge $c$).
 The Luttinger liquid particle number operator $N_\el=-\sqrt{\nu_H}J_0$ is expressed in 
terms of the zero mode $J_0$ of the normalized $\uu$ current and 
$\Delta \varepsilon=\hbar 2\pi v_c/L$ is the non-interacting energy spacing in the QD. 
The Hilbert space $\H$ of the FQH edge states, over which the trace is taken,
 depends on the type and number of the localized FQH quasiparticles in the bulk.

 When the magnetic field $B$ or the area $A$ or the gate voltage $V_g$ are changed from their initial values, $B_0$, 
or $A_0$ respectively,  the partition function $Z(\beta,\mu)=Z(\t,\z)$ is modified by shifting the modular parameters, 
as proven in \cite{NPB-PF_k} (see Eq.~(34))
\beq\label{shift}
\z\to \z+\phi\t, \quad \phi = \frac{C_g}{e\nu_H}V_g ,
\eeq
where the modular parameters $\t$ and $\z$, used to construct (rational) CFT partition functions  \cite{CFT-book},
are related to the temperature $T$ and chemical potential $\mu$ by
$\t=i\pi T_0/T$,  $T_0=\hbar v_c/\pi k_B L$, $\z=(\mu /\Delta\varepsilon)\t$.
To understand Eq.~(\ref{shift}) physically we recall the Aharonov--Bohm relation: 
the electron field operator $\psi_{\mathrm{el}}(z)$, where $z=\e^{i\varphi}$ is 
the electron coordinate on the edge circle, is modified 
in presence of AB vector potential $\bf{A}$ as $\psi^{\bf A}_{\mathrm{el}}(z)=z^{-\phi}\psi_{\mathrm{el}}(z)$, where 
$\phi$ 
is the dimensionless AB flux (see Eq.~(26) in \cite{NPB-PF_k}). The AB flux changes the boundary conditions of all charged 
particles operators
and the adiabatic variation of the flux changes the Hilbert space of the edge excitations by a well known procedure called 
\textit{twisting} in the conformal field theory \cite{CFT-book,kt,NPB-PF_k}.
In particular, the partition function (\ref{Z0})  changes as \cite{NPB-PF_k} 
\beq \label{Z2}
Z_{\phi}(\beta,\mu)= \mathop{\mathrm{tr}}_{\mbox{\quad}\H} \e^{-\beta (H_{\mathrm{CFT}}(\phi) -\mu N_{\imb})},
\eeq
where $\H$ is the untwisted Hilbert space, corresponding to $\phi=0$, the thermodynamic parameters $\beta$ and $\mu$ are
 independent of $\phi$, and all flux dependence is moved to the twisted
operators of energy $H_{\mathrm{CFT}}(\phi)$ and charge imbalance\cite{staring-CB} $N_{\imb}$ 
(cf. Eqs.~(32) and (33) in \cite{NPB-PF_k})
\beq\label{H-N}
H_{\mathrm{CFT}}(\phi)=H_{\mathrm{CFT}} -\Delta\varepsilon \phi N_\el +\frac{\nu_H}{2} \Delta\varepsilon\phi^2, \quad 
N_{\imb}=N_\el -\nu_H \phi .
\eeq
The ultimate effect of the AB flux on the partition function $Z(\beta,\mu)$  is shifting $\zeta$ as in (\ref{shift}) or, equivalently,
$\mu \to \mu+\phi\Delta \varepsilon$.

It follows from (\ref{Z2}) that $\partial \Omega / \partial \mu = -\la N_{\imb} \ra_{\beta,\mu}$ and taking into account (\ref{H-N})
we find that the thermodynamic average of the electron number in presence of AB flux (setting $\mu=0$)  is
\beq \label{N}
\la N_\el (\phi)\ra_{\beta,\mu} = -\frac{\partial \Omega_{\phi}(T,\mu)}{\partial \mu} +\nu_H \phi ,
\eeq
where $\Omega_{\phi}(T,\mu)=-k_B T \ln Z_{\phi}(\beta,\mu)$.
This general construction of the electron number operator average in presence of AB flux allows us to compute also the 
flux dependence of the conductance of the Coulomb island according to Eq.~(10) in \cite{thermal} and 
Eq.~(\ref{N}) at $\mu=0$, i.e., 
\beq \label{G}
G (\phi)=\frac{e^2}{h}
\left( \nu_H +\frac{1}{2\pi^2} \left(\frac{T}{T_0} \right)\frac{\partial^2 }{\partial \phi^2}  \ln Z_{\phi}(T,0)\right) .
\eeq
The electron number (\ref{N}) and the conductance (\ref{G}) are illustrated for the Pfaffian FQH state in Fig.~\ref{fig:G-N} below.
Next, we can compute the average quantum dot energies with $N$ electrons on the edge at temperature $T$ and 
chemical potential $\mu$ in presence of AB flux from the standard Grand 
canonical ensemble relation \cite{kubo}
\beq \label{H_CFT}
\la H_{\mathrm{CFT}}(\phi)\ra_{\beta,\mu} =\Omega_{\phi}(\beta,\mu)- T \frac{\partial \Omega_{\phi}(\beta,\mu)}{\partial T}  
 - \mu\frac{\partial  \Omega_{\phi}(\beta,\mu)}{\partial \mu}.
\eeq
To summarize the main result in this paper: substituting Eq.~(\ref{H_CFT}) and (\ref{N}) into Eq.~(\ref{eps}) we can compute the thermopower of a CB 
FQH island in terms of the edge state's partition function (\ref{Z2}) in presence of Aharonov--Bohm flux or side-gate voltage 
introduced into the latter through Eq.~(\ref{shift}).
\section{Thermopower for general FQH disks}
\label{sec:disk}
The Grand canonical partition function (\ref{Z0}) for a general FQH  edge states on a disk can be written 
as \cite{cz,LT9,cappelli-viola-zemba,cappelli-viola}
\beq \label{Z-gen}
Z^{l,\Lambda}(\t,\z)=\sum_{s=0}^{n_H-1} K_{l+s d_H}(\t,n_H\z; n_H d_H) \ch_{\omega^s * \Lambda}(\t'), \quad
\eeq
where $n_H$ and $d_H$ are the numerator and denominator of the filling factor $\nu_H=n_H/d_H$ while $\omega$ is the neutral 
topological charge of the electron operator, which is always non-trivial \cite{NPB-PF_k} when $n_H>1$. 
The $\uu$ partition function $K_{l}(\t,\z; m)$ for the charged part is completely determined by the filling factor $\nu_H$ and 
coincides with that for a chiral Luttinger liquid with a compactification radius \cite{cz} $R_c=1/m$, in the 
notation of  \cite{CFT-book,NPB-PF_k} 
 \beq \label{K}
K_{l}(\t,\z; m) = \frac{\mathrm{CZ}}{\eta(\t)} \sum_{n=-\infty}^{\infty} q^{\frac{m}{2}\left(n+\frac{l}{m}\right)^2} 
\e^{2\pi i \z \left(n+\frac{l}{m}\right)},
\eeq
where  $q=\e^{2\pi i \t }$, 
$\eta(\t)=q^{1/24}\prod_{n=1}^\infty (1-q^n)$ is the Dedekind function \cite{CFT-book} and 
 $\mathrm{CZ}(\t,\z)=\exp(-\pi\nu_H(\Im \z)^2/\Im\t)$ is the Cappelli--Zemba factor  introduced to restore the 
invariance of $K_{l}(\t,\z; m)$ with respect to the Laughlin spectral flow \cite{cz}.
The $\uu$ charge label $l$ in $Z^{l,\Lambda}$ is determined by the 
total electric charge of the localized quasiparticles in the bulk   $Q_{\mathrm{el}}(\mathrm{bulk}) = l/d_H$,  while the 
weight $\Lambda$
is determined by the total neutral topological charge of the quasiparticles localized in the bulk. The partition 
function $\ch_{\Lambda}(\t')$ 
represents the neutral edge modes corresponding to the total neutral topological charge $\Lambda$ in the bulk. 
The modular parameter 
$\t'=r \t$ with $r= v_n/v_c$ is modified in order to take into account the observation \cite{neutral-velocity,viola-stern} that the 
Fermi velocity $v_n$ of the neutral edge modes might be smaller than $v_c$. 
 The $*$ in $Z^{l,\Lambda}$  denotes the fusion product \cite{CFT-book} of the topological charge $\Lambda$ with the 
($s$-multiple) neutral topological charge $\omega$ of the electron \cite{NPB-PF_k,LT9}.
The electric charge of the edge excitations, with quantum numbers $n$ and $s$, encoded in the partition function (\ref{Z-gen})  is
\beq \label{Q-gen}
Q_\el(l, \Lambda,n,s)=\frac{l}{d_H}+s +n_H n,
\eeq
where $l$ is the number of fundamental quasiparticles in the bulk, $s$ is the number of electrons on the edge and $n$ is the number
of clusters of  $n_H$ electron on the edge.
The neutral topological weight $\Lambda$ and the electric charge $l$ have to satisfy a general 
$\Z_{n_H}$ pairing rule, see Eq.~(19) in \cite{LT9},
\beq \label{PR}
n_H \tilde{Q}_{\omega}(\Lambda)  \equiv l \mod n_H.
\eeq
where $\tilde{Q}_{\omega}(\Lambda)$ is the (neutral) monodromy charge defined by the following combination of 
conformal dimensions $\D_{\Lambda}$ of the neutral Virasoro irreducible representations
\beq \label{monodromy}
 \tilde{Q}_{\omega}(\Lambda) \equiv \D_{\omega*\Lambda} -\D_\Lambda -\D_\omega \mod \Z, 
\quad \left(\D_\omega = \D^{(0)} \right).
\eeq
The $\Z_{n_H}$ pairing rule (\ref{PR}), which selects the admissible pairs $(l,\Lambda)$ of charged and neutral quantum numbers,
follows from the the locality condition for the short-distance operator product expansions of the physical excitations 
with respect to the neutral part $\psi_\el^{(0)}(z)$ of the electron field \cite{LT9} of CFT dimension $\Delta^{(0)}$, 
characterized by the neutral electron weight $\omega$.

Next we can introduce the AB flux $\phi$ into the partition function (\ref{Z-gen}) by the shift (\ref{shift}) and then move the flux and chemical 
potential dependences into the charge index of the $K$ function (\ref{K}), due to the identity  \cite{NPB-PF_k}
\beq \label{K-shift}
K_{l}(\t,\z+\phi\t; m)\equiv K_{l+\phi}(\t,\z; m),
\eeq
setting $\z=0$ after that, i.e. introducing AB flux $\phi$ leads to the index shift $l+sd_H\to l+sd_H +n_H \phi$, 
so that the partition function with AB flux $\phi$ is\footnote{we skip the $\eta$-function and the CZ factor from Eq.~(\ref{Z-gen})
which are unimportant multiplicative factors for $\z=0$}
\beq\label{Z_l-L-phi}
Z_{\phi}^{l,\Lambda}(\t,{\mu})\propto \sum_{s=0}^{n_H-1}  \ch_{\omega^s * \Lambda}(\t')  
  \sum_{n=-\infty}^{\infty} \e^{-\beta\Delta \varepsilon\frac{n_Hd_H}{2}\left(n+\frac{l+sd_H}{n_H d_H}+
\frac{1}{d_H}\left( \phi + \frac{\mu}{\Delta\varepsilon}\right)\right)^2}. \quad 
\eeq
In order to compute the average tunneling energy (\ref{eps}) we need to specify the parameters $\mu_{N}$ and 
$\mu_{N+1}$ corresponding to QD with $N$ and $N+1$ electrons, respectively. To this end we emphasize that 
the parameter $\mu$ entering (\ref{Z2}) is not the true chemical potential, except at zero gate voltage, because 
it is not coupled to the electron number operator
but to the charge imbalance $N_{\imb}$ \cite{staring-CB}.

 As can be seen in \ref{app:mu}, 
when the bulk  electron number is $N_0=n_H n_0$, where  $n_0$ is a positive integer and $n_H$ is the numerator 
of the filling factor $\nu_H$, the partition function (\ref{Z_l-L-phi}) 
is independent of the  bulk component of $\mu$ and the edge components of $\mu_{N}$ and $\mu_{N+1}$ can be chosen as
\beq \label{mu_NN1}
\mu_N=-\frac{\Delta \varepsilon}{2} \quad \mathrm{and} \quad \mu_{N+1}=\frac{\Delta \varepsilon}{2},
\eeq
where $\Delta \varepsilon$ is the QD level spacing. This choice of $\mu_N$ and $\mu_{N+1}$ is universal in the sense that it is
independent of the neutral contributions to the electron energy, or of the ratio $r=v_n/v_c$ of the Fermi velocities of the neutral and 
charged modes, or even of  $\nu_H$, see Appendix~\ref{app:mu} for more detail.
\section{Thermoelectric properties of a $\nu_H=5/2$ CB island}
\label{sec:5-2}
We will illustrate the general approach to thermopower using as an example a $\nu_H=5/2$ CB island, assuming that only the higher 
$\nu_H=1/2$ edge is strongly backscattering, as in Ref.~\cite{viola-stern}. So far only the fractional electric 
charge $e/4$ of the fundamental quasiparticles has been confirmed experimentally 
\cite{dolev-charge-1-4,radu-charge-1-4,willet-charge1-4} and this is consistent with several FQH candidates: 
the Pfaffian Moore--Read state \cite{mr}, the anti-Pfaffian \cite{rosenow-APf,nayak-APf}, the 331 Halperin state \cite{331-halperin}
and the $\uu \times \widehat{SU(2)}_2$ state \cite{overbosch-wen}. The difference between the Pfaffian and the 
$\widehat{SU(2)}_2$ state is 
only in the neutral sector: instead of the Ising model irreducible representations with CFT dimensions $0$, $1/16$ and $1/2$ we 
use in the $\widehat{SU(2)}_2$ state  the irreducible representation of the level-2
current algebra $\widehat{SU(2)}_2$  with CFT dimensions $0$, $3/16$ and $1/2$, respectively. The electric properties of the 
Pfaffian and $\widehat{SU(2)}_2$ are the same, the fusion rules are the same, the CB peak patterns are the same in the zero 
temperature limit, yet they have significantly different thermoelectric properties which if measured in an experiment could 
help us to figure out which state is realized in nature. 
For our purposes, the most important things are the partition functions for the two models and they are explicitly 
written in Eqs.~(\ref{ch_Pf}) and (\ref{chi}) below.

In the rest of this paper we will use the CFT partition functions to
compute numerically the thermopower, conductance, thermal conductance and power factor 
for  these FQH islands. 
The partition functions of the paired FQH states with $\nu_H=5/2$ with quasiparticles in the bulk depend on their number modulo 2
(modulo $n_H$ in general) \cite{stern-CB-RR,cappelli-viola-zemba,viola-stern}. Therefore we consider only two cases: even 
(no quasiparticles in the bulk) and odd (one quasiparticle in the bulk). We start with the case of one bulk quasiparticle (odd number),
 which is simpler than the case of zero bulk quasiparticles (even number).
\subsection{Odd number of bulk quasiparticles}
The partition functions for the 
paired\footnote{we consider only the highest Landau level with $\nu_H=1/2$ and $n_H=2$, $d_H=4$}  
$\nu_H=1/2$ FQH states with odd number of quasiparticles in the bulk, e.g. 
with one quasiparticle in the bulk, is 
\beqa\label{Z-od}
Z^{\mathrm{odd}}(\tau,\zeta)&=&K_1(\t,2\z;8)\ch_\s(\t') + K_{-3}(\t,2\z;8)\ch_{\omega*\s}(\t')\nn
&=&\left[K_1(\t,2\z;8)+K_{-3}(\t,2\z;8)\right]\ch_\s(\t'),
\eeqa
where the $K$ functions are defined in (\ref{K}), $\ch_\s(\t')$ is the neutral partition function for the one-quasiparticle sector
labeled by the topological charge $\s$ of the basic quasihole in the FQH liquid, which could generate by fusion with itself 
all other quasiholes, and $\ch_{\omega*\s}$ is the neutral partition function of the topological sector labeled by
$\omega*\s$, with $\omega$ being the neutral topological charge of the electron filed \cite{NPB-PF_k,LT9}. 
In all three cases of the Pfaffian, $SU(2)_2$ and Anti-Pfaffian the neutral characters 
satisfy $\ch_{\omega*\s}(\t')\equiv \ch_{\s}(\t')$ as a consequence of the fusion rules $\psi_\el^{(0)}\times \s \simeq \s$, where $\s$ 
is the lowest-CFT dimension quasiparticle field and $\psi_\el^{(0)}$ is the electron field corresponding to the neutral 
topological weight $\omega$. For the 331 state the sectors with $\s$ and $\omega*\s$ are represented by different topological charges
having opposite fermion parity,
however the neutral partition functions of the two sectors coincide because $K_{-1}(\t,0;4)\equiv K_{1}(\t,0;4)$. Therefore in all four cases
of paired FQH states, including the Anti-Pfaffian state, we have $\ch_{\omega*\s}(\t')\equiv \ch_{\s}(\t')$ which explains the second 
line of (\ref{Z-od}). Next we can use the identity \footnote{the relation (\ref{K-sum}) is not so obvious although it follows directly from 
Eq.~(\ref{K}). It is the odd-index version of a more general relation of Eq. (C.5) in Ref. ~\cite{CMP99}.}
\beq \label{K-sum}
K_{l+\frac{1}{2}}(\tau,\zeta;2) \equiv K_{2l+1}(\tau,2\zeta;8)+K_{2l-3}(\tau,2\zeta;8)
\eeq
to find that
$K_1(\t,2\z;8)+K_{-3}(\t,2\z;8)= K_{1/2}(\t,\z;2)$ and the partition function (\ref{Z-od}) is written as \textit{a single product} 
of a charged part $K_{1/2}(\t,\z;2)$  and a neutral part $\ch_{\s}(\t')$. It is now obvious that the neutral part of the CFT has no 
contribution to the average tunneling energy (4) for odd number of quasiparticles in the bulk and the distance between the 
centers of the consecutive peaks of the electric and thermal conductances is always $\Delta\phi =2$, as shown in 
Fig.~\ref{fig:G-GT-Pf-odd-T1} below. This result is consistent with that of Ref.~\cite{viola-stern}. 
 Therefore, the neutral degrees of freedom are completely decoupled from the charged ones and the 
thermoelectric properties are basically the same as for the $g=1/2$ Luttinger liquid. 
This includes the anti-Pfaffian state, with odd number of bulk quasiparticles,  as well, cf. \cite{viola-stern}. Thus,
the thermoelectric quantities of the paired states with odd number of bulk quasiparticles are not sensitive to the ratio 
$r=v_n/v_c$ of the Fermi velocities of the neutral and charged edge modes.

\begin{figure}[htb]
\centering
\includegraphics[bb=40 15 580 375,clip,width=9cm]{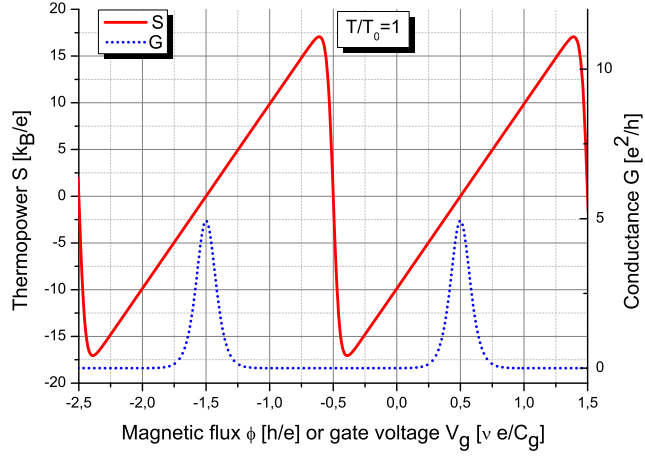}
\caption{Thermopower of a CB island for all paired states for $\nu_H=5/2$, with odd number of quasiparticles in the bulk,
 including anti-Pfaffian, at temperature $T/T_0=1$. \label{fig:TP-G-Pf-odd-T10-r1}}
\end{figure}

The thermopower for a Coulomb blockaded island,  of paired $\nu_H=5/2$ quantum Hall states with one quasiparticle in the bulk,
as a function of the AB flux is computed numerically from Eqs.~(\ref{S}), (\ref{eps}), (\ref{N}), (\ref{H_CFT}) and (\ref{Z_l-L-phi}) and
 is given graphically in Fig.~\ref{fig:TP-G-Pf-odd-T10-r1}.
The thermopower oscillations are apparently similar to those of metallic islands except that the period in the normalized AB flux
 is $\Delta\phi =2$ instead of $1$. The peaks of the electrical conductance shown in Fig.~\ref{fig:TP-G-Pf-odd-T10-r1} are equally 
spaced with period $\Delta\phi =2$, the thermopower is zero at the centers of the conductance peaks and thermopower is 
disconnected (at $T=0$) at the centers of the CB valleys, just like in metallic CB islands \cite{matveev-LNP,matveev-PRB}.
For comparison, the oscillation of the thermopower, for a  CB island formed in a $\nu_H=1/3$ 
Laughlin quantum Hall state with $l=0$ and the CB conductance, can be seen in \cite{LT10}. 
The thermopower in that case has a similar saw-tooth behavior with period 3 flux quanta and  
the positions of the CB peaks correspond to the zeros of the thermopower.
cf.  \cite{matveev-LNP,matveev-PRB}.  
 
\begin{figure}[htb]
\centering
\includegraphics[bb=60 25 530 270,clip,width=8.5cm]{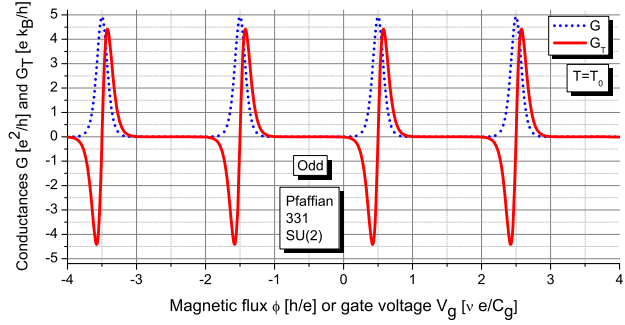}
\caption{Electric and thermal conductances $G$ and $G_T=S.G$  for all paired FQH states with odd number of quasiparticles 
in the bulk, including the Anti-Pfaffian CB island at temperature $T/T_0=1$. \label{fig:G-GT-Pf-odd-T1}}
\end{figure}
Due to Eq.~(\ref{S-G}) the thermal conductances peaks for all paired FQH states with odd number of bulk quaiparticles 
can be obtained from the calculated thermopower $S$ and electrical conductance (\ref{G}) by $G_T=G.S$, see
Fig.~\ref{fig:G-GT-Pf-odd-T1}.
They are also equally spaced, with period $\Delta \phi =2$, are completely symmetric for all temperatures
and are independent of $r=v_n/v_c$, showing similar results as in \cite{viola-stern}.

\begin{figure}[htb]
\centering
\includegraphics[bb=30 10 570 380,clip,width=8.5cm]{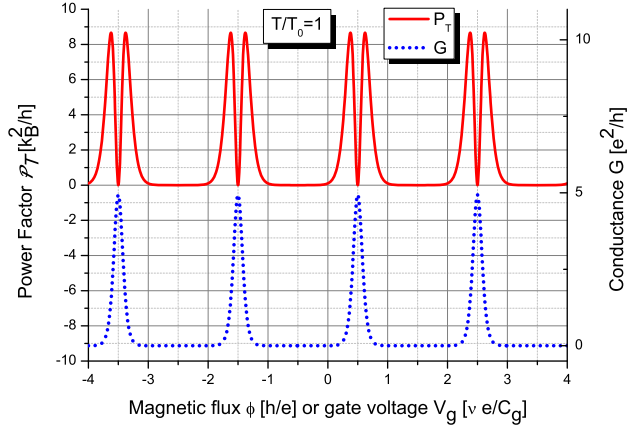}
\caption{The Power factor $\P_T$ of a CB island for all paired states for $\nu_H=5/2$, with odd number of quasiparticles in the bulk,
 including anti-Pfaffian, at temperature $T/T_0=1$. \label{fig:PF-Pf-T10}}
\end{figure}

Finally, in Fig.~\ref{fig:PF-Pf-T10} we plot the power factor $\P_T$ for all paired states at $\nu_H=5/2$ with odd number of 
bulk quasiparticles computed numerically at $T=T_0$ from Eq.~(\ref{P_T}), together with the peaks of the conductance 
(right-Y scale). 
The power factor shows sharp dips corresponding precisely to the maxima of the conductance peaks. 
Notice that this figure is qualitatively similar to Fig.~3c in Ref.~\cite{gurman-2-3}, 
which suggests that the method used there for measuring the thermoelectric current might be convenient for measuring the 
power factors of for the $\nu_H=5/2$ FQH state as well, see the discussion before Eq.~(\ref{power}).
\subsection{Even number of bulk quasiparticles}
The partition functions for a CB island in all paired FQH states with $\nu_H=5/2$  
with even number of quasiparticles in the bulk 
\footnote{we consider only the highest Landau level with $\nu_H=1/2$ and $n_H=2$, $d_H=4$} 
can be written as  a sum of two products,  
e.g. for zero bulk quasiparticles,
\beq\label{Z-ev}
Z^{\mathrm{even}}(\tau,\zeta)=K_0(\t,2\z;8)\ch_0(\t') + K_4(\t,2\z;8)\ch_{\omega}(\t')
\eeq
where $\t' = r\t$ with $r=v_n/v_c$, the $K$ functions are defined in (\ref{K}) and  $\ch_0(\t')$ is the neutral partition function 
of the vacuum sector, while $\ch_\omega(\t') $ is the neutral partition function
of the  one-electron sector. 

\begin{figure}[htb]
\centering
\includegraphics[bb=40 10 530 380,clip,width=8.5cm]{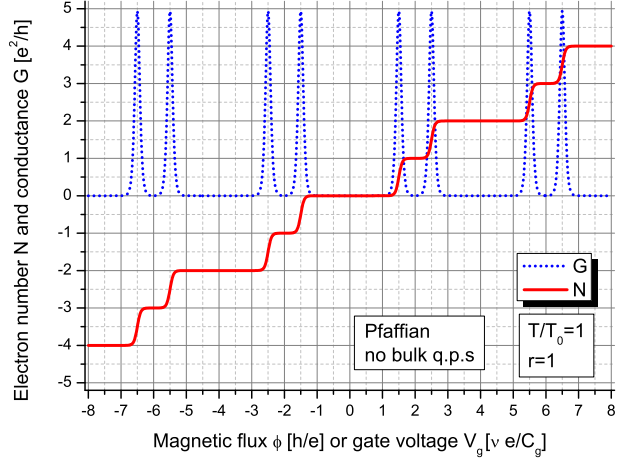}
\caption{Coulomb island conductance  $G$ and electron number $N$, measured from $N_0$, of the 
Pfaffian FQH state at temperature $T=T_0$ for $v_n=v_c$. The number $N$ jumps by $1$ at the 
conductance peak positions.  \label{fig:G-N}}
\end{figure}

The neutral partition functions in (\ref{Z-ev}) for the Pfaffian state can be expressed as
\beqa \label{ch_Pf}
\ch_{0}(\t)&=& \frac{q^{-1/48}}{2}\left(\prod_{n=1}^\infty (1+q^{n-1/2}) + \prod_{n=1}^\infty (1-q^{n-1/2})\right), \nn
\ch_{\omega}(\t)&=& \frac{q^{-1/48}}{2}\left(\prod_{n=1}^\infty (1+q^{n-1/2}) -  \prod_{n=1}^\infty (1-q^{n-1/2})\right),
\eeqa 
where $q=e^{2\pi i \t}$. Now that we have specified the complete partition function for the Pfaffian state without bulk 
quasiparticles we plot in Fig.~\ref{fig:G-N} the electron number (\ref{N}) and electric 
conductance (\ref{G})  for the Pfaffian island without bulk quasiparticles as functions of the AB flux $\phi$, 
respectively, of the gate voltage $V_g$.
We see in Fig.~\ref{fig:G-N} that the positions of the peaks of the electric conductance of the CB island precisely 
corresponds to the positions in gate voltage where the electron number on the island increases by one. For illustration
purposes the two functions are computed for equal velocities of the neutral and charged edge modes, $v_n=v_c$, i.e. for 
$r=1$ in which case the conductance peaks are packed in pairs of peaks separated by flux distance $\Delta\phi_1=1$ which
are then separated by a flux distance $\Delta\phi_2=2$ between the groups of peaks. When temperature is decreased 
the CB conductance peaks become higher and narrower while the profile of the electron number becomes closer to the 
step function.

\begin{figure}[htb]
\centering
\includegraphics[bb=30 0 580 385,clip,width=10.5cm]{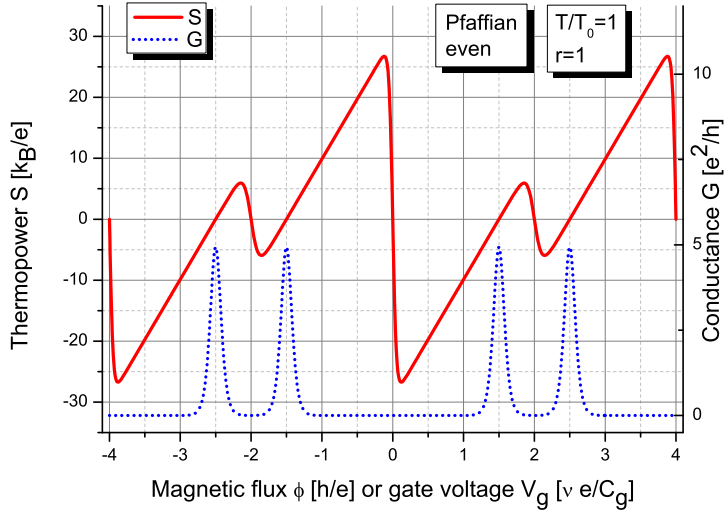}
\caption{Thermopower of a CB island in the Pfaffian state for $\nu_H=5/2$, without quasiparticles in the bulk, for $r=1$,
at temperature $T/T_0=1$. \label{fig:TP-Pf-T10-r1}}
\end{figure}
The thermopower for a Coulomb blockaded island in the Pfaffian state without bulk quasiparticles with $r=1$ 
is computed from Eqs.~(\ref{S}), (\ref{eps}), (\ref{N}), (\ref{H_CFT}) and (\ref{Z_l-L-phi}) and is given as a function of the gate voltage in 
Fig.~\ref{fig:TP-Pf-T10-r1}, which is a central result in this work.
Two important characteristics of the thermopower for fractional quantum Hall states  have to be emphasized:
when the gate voltage approaches a position of a CB peak the thermopower vanishes at the maximum of the peak, just like it 
does for metallic islands \cite{matveev-LNP}; second, at the centers of the CB valleys thermopower decreases rapidly 
(jumping discontinuously at $T=0$) crossing the $x$-axis exactly at the center of the valley like in metallic islands 
\cite{matveev-LNP}. This modified saw-tooth shape of the thermopower is similar to that in 
superconducting SET \cite{TP-SC}.

\begin{figure}[htb]
\centering
\includegraphics[bb=40 15 580 385,clip,width=10cm]{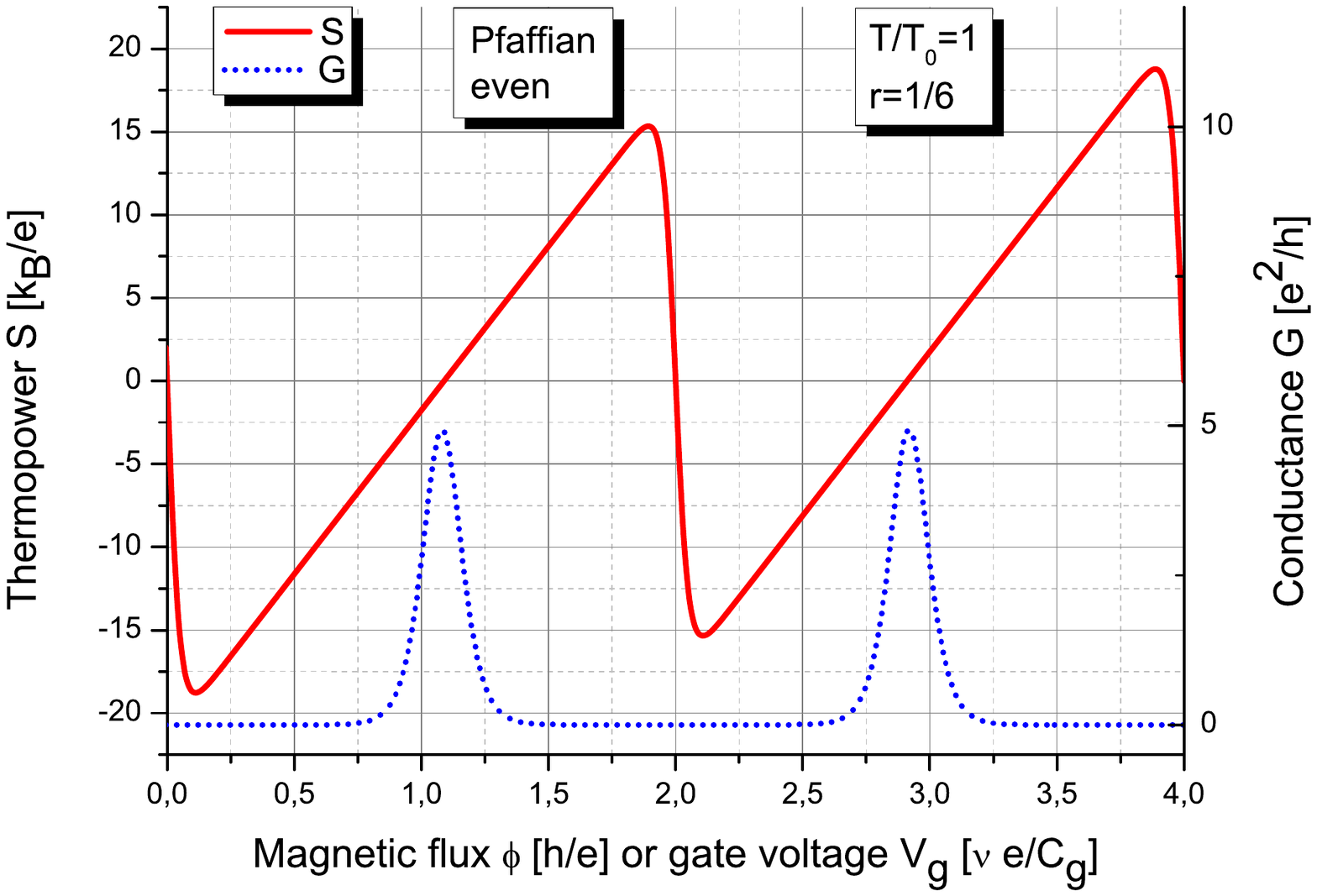}
\caption{Thermopower of a CB island in the Pfaffian state for $\nu_H=5/2$, without quasiparticles in the bulk, for $r=1/6$,
at temperature $T/T_0=1$. \label{fig:TP-Pf-T10-r1-6}}
\end{figure}
In addition to Fig.~\ref{fig:TP-Pf-T10-r1} showing the thermopower for the Pfaffian state with even number of bulk quasiparticles 
at $T=T_0$ and $r=1$, we plot for comparison in Fig.~\ref{fig:TP-Pf-T10-r1-6} the thermopower for the same state (Pfaffian, even),
however with $r=1/6$.
We see that the differences in the neighboring maxima of the thermopower, seen in Fig.~\ref{fig:TP-Pf-T10-r1}, 
decreased for $r=1/6$ and are 
probably hard to observe experimentally. However, still the zero's of the thermopower correspond to the maxima of the electric 
conductance. Again the CB peaks are not equally spaced and have two periods which are of the form $\Delta \phi_1=2-r$ and
 $\Delta \phi_2=2+r$, where $r=v_n/v_c$ is the ratio of the Fermi velocities of the neutral and charged edge modes. However, 
measuring the differences in CB peak spacing does not seem very promising from the experimental point of view.

The plot of the thermopower of the Anti-Pfaffian state is similar to that in Fig.~\ref{fig:TP-Pf-T10-r1-6} except that the positions of 
the higher and lower maxima of the thermopower are exchanged.

\begin{figure}[htb]
\centering
\includegraphics[bb=10 5 580 795,clip,width=8.5cm]{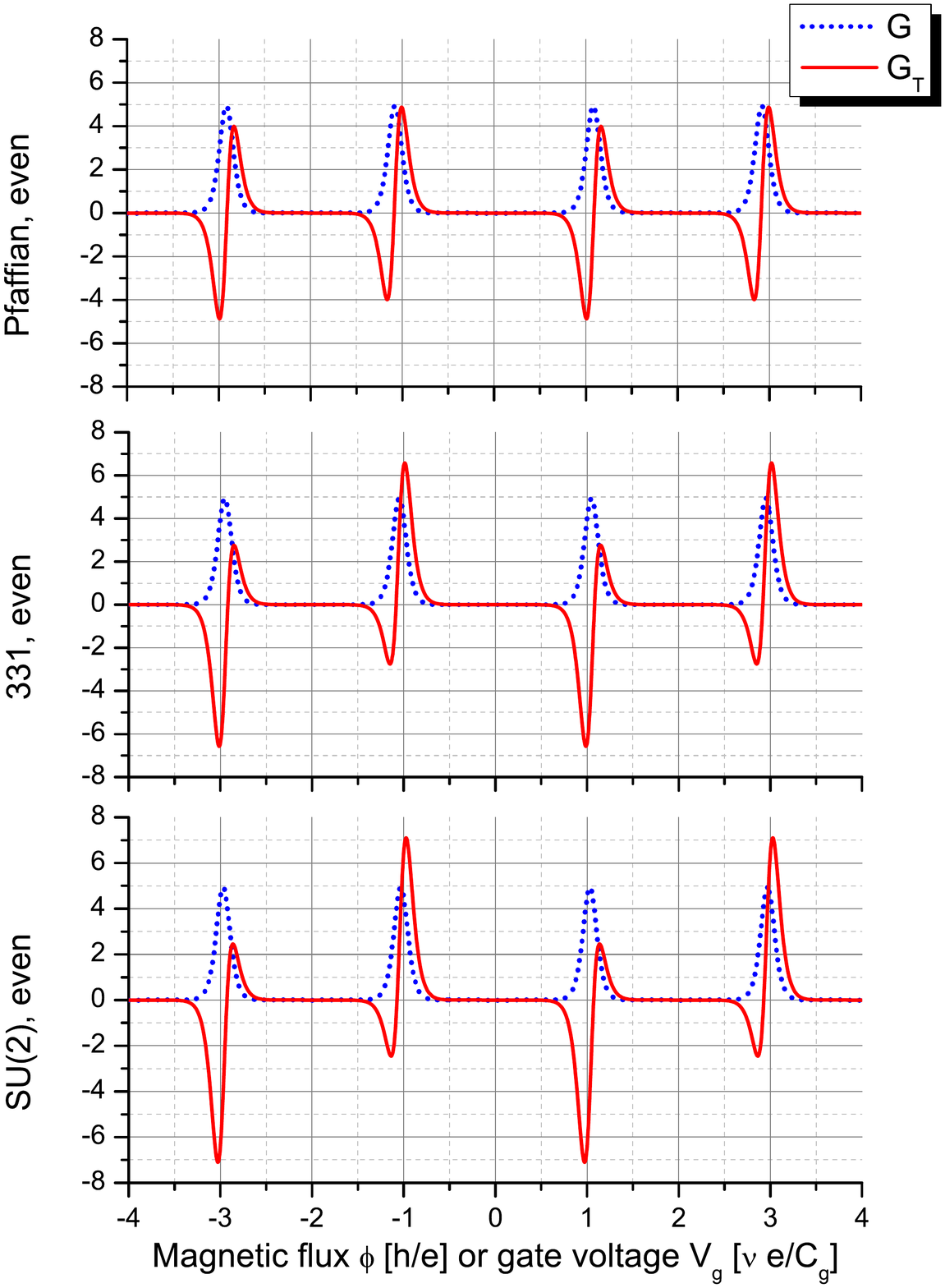}
\caption{Conductances $G$, in units [$e^2/h$], and $G_T$, in units [$ek_B/h$], for the Pfaffian, 331 and $SU(2)_2$ 
models without quasiparticles in the bulk, with $r=1/6$, at temperature $T/T_0=1$. \label{fig:G-GTs}}
\end{figure}

Next, we continue with the other paired FQH states. In order to completely define the corresponding total partition functions 
(\ref{Z-ev}) for the edges of the CB island we need to specify the neutral partition functions.
The neutral partition functions for a Coulomb blockaded island in the 331 state are expressed in terms of the $K$ functions (\ref{K})  
as 
\beq \label{ch_331}
\ch_0(q)=K_0(\tau,0;4)  , \quad \ch_\omega(q)=K_2(\tau,0;4), 
\eeq
while the neutral partition functions for the $SU(2)_2$  FQH state \cite{overbosch-wen} are defined as the 
$\widehat{SU(2)}_2$ characters with lowest CFT dimensions $0$ and $1/2$ respectively, which are expressed in terms
of the functions from Eq.(14.183) in ~Ref.~\cite{CFT-book}
\beqa
\ch_{0}(\t)=\chi_{0}^{(2)}(\t)&=&\frac{q^\frac{1}{16}}{\eta(q)^3}\sum_{n\in\Z}(1+8n)q^{n(1+4n)}\nn
\ch_{\omega}(\t)=\chi_{2}^{(2)}(\t)&=&\frac{q^\frac{9}{16}}{\eta(q)^3}\sum_{n\in\Z}(3+8n)q^{n(3+4n)}.\label{chi}
\eeqa
In Fig.~\ref{fig:G-GTs} we plotted the electric and thermal conductances for the Pfaffian, 331 and $SU(2)_2$ states
with even number of  quasiparticles in the bulk at $T=T_0$ for $r=v_n/v_c =1/6$.
We see that the zeros of the thermal conductances $G_T$ coincide with the maxima of the corresponding 
electric conductances' peaks and then $G_T$ changes sign. However, unlike the case with odd number of 
bulk quasiparticels, where the peaks were equally spaced, here there are again the two periods 
$\Delta \phi_1=2-r$ and $\Delta \phi_2=2+r$, depending on the  ratio $r=v_n/v_c$ of the Fermi velocities of the neutral and 
charged edge modes.
The oscillations of the thermal conductances in Fig.~\ref{fig:G-GTs}  are apparently asymmetric which is a consequence of
the asymmetries in the thermopower. These asymmetries are signals that the 
neutral degrees of freedom play an important role for the paired FQH states with even number of bulk 
quasiparticles \cite{thermal,viola-stern}. 

As pointed out in Ref.~\cite{viola-stern} the ratio of the amplitudes of the minimum and maximum of the thermal conductance $G_T$
could serve as an experimental signature that could eventually distinguish between the different paired FQH states having the same 
electric conductance peak patterns. Below we will demonstrate that the ratio of the neighboring maxima of the 
power factor (\ref{P_T}) around a CB peak position, computed from the thermopower and the conductances, 
might be a better tool to distinguish between these FQH states.  We argue that this quantity could be measured by the method 
of Ref.~\cite{gurman-2-3} where the N-gate input AC voltage $V_N$ has  frequency $f_0/2=497.5$ kHz,
while the transmission coefficient is measured for the output current at frequency $f_0=995$ kHz. We claim that the measured signal
shown in Fig. 3c in Ref.~\cite{gurman-2-3} is actually proportional the the power factor (\ref{P_T}). Indeed, the measured current
is of the form $I= G \Delta V +G_T \Delta T$, where $\Delta V$ is the QD bias  which we assume to be very small and
the voltage induced temperature difference is $\Delta T =\alpha |V_N|$ for small $\Delta T$, as can be seen in 
Fig.~2c in Ref.~\cite{gurman-2-3}. The voltage on gate N in the setup of 
 Ref.~\cite{gurman-2-3} is of the form $V_N=V_{N,DC} + V_{N,AC} \cos(2\pi (f_0/2) t)$ and in the regime where the input frequency
 is half of that of the measured output signal $V_{N,DC}$ is set to $0$. When the input signal $V_N$ is at frequency $f_0/2$
 and the output is measured at $f_0$ we are actually measuring the square of the current, which is 
$I^2 \propto V_{N,AC}^2/2 \cos(2\pi f_0 t)$. On the other hand, if we consider the power of the thermoelectric 
current\footnote{we assume that the bias is $\Delta V\simeq 0$}
\beq \label{power}
P=G I^2 \propto \alpha^2 G. S^2 V_N^2 \propto P_T \cos(2\pi f_0 t),
\eeq
we see that the term in front of $\cos(2\pi f_0 t)$ is proportional to the power factor $P_T$ defined in (\ref{P_T}).
Thus we conclude that the signal shown in Fig. 3c in Ref.~\cite{gurman-2-3} is proportional to the 
power factor (\ref{P_T}) for $\nu_H=2/3$ which is similar to our Fig.~\ref{fig:PF-Pf-T10}.
Therefore we are confident that this procedure gives a nice method for direct measuring the power factor of a QD.
\begin{figure}[htb]
\centering
\includegraphics[bb=25 70 550 770,clip,width=7.5cm]{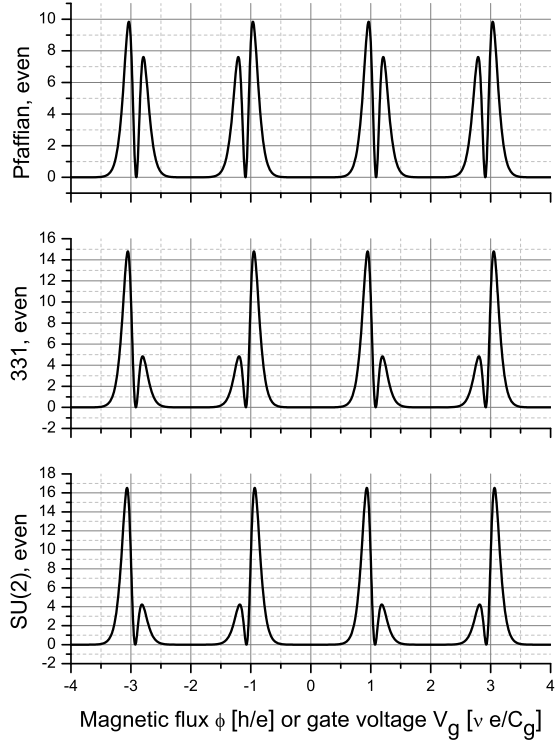}
\caption{Power factors $\P_T$, in units [$k_B^2/h$], for the Pfaffian, 331 and $SU(2)_2$ models without quasiparticles 
in the bulk, with $r=1/6$, at temperature $T/T_0=1$. \label{fig:PFs}}
\end{figure}
Next, we show in Fig.~\ref{fig:PFs} the power factors for the Pfaffian, 331 and $SU(2)_2$ models without quasiparticles 
in the bulk, with $r=1/6$, at temperature $T/T_0=1$.
We emphasize here that, due to the asymmetries mentioned earlier, the plots of the power factors for the different 
paired states candidates describing the $\nu_H=5/2$
FQH state are noticeably different for the different FQH states even for $r=1/6$ when the electric conductance patterns are 
indistinguishable.
Therefore we strongly believe that the power factor (\ref{P_T}) is a better spectroscopic tool than the electric and thermal 
conductances alone, which might eventually be used to experimentally distinguish between the different candidate states 
for $\nu_H=5/2$.

\begin{figure}[htb]
\centering
\includegraphics[bb=30 10 570 390,clip,width=10.5cm]{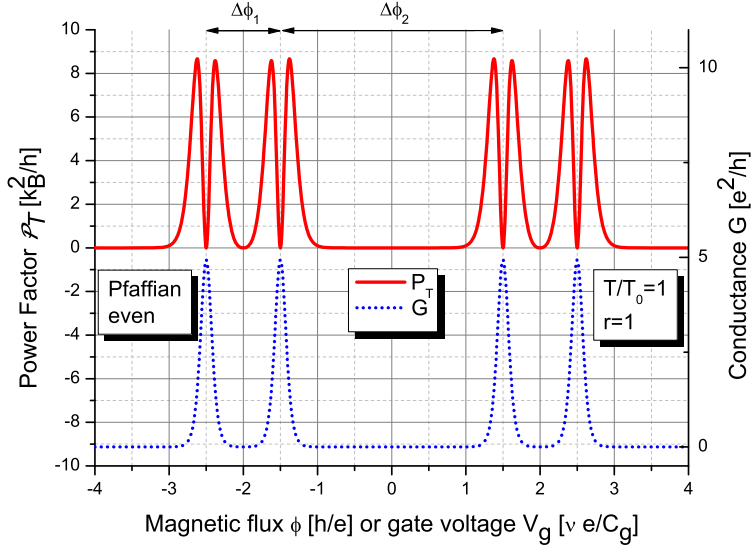}
\caption{The Power factor $\P_T$ of a CB island for the Pfaffian state for $\nu_H=5/2$, without quasiparticles in the bulk,
 with $r=1$, at temperature $T/T_0=1$. \label{fig:PF-G-Pf-even0T10-r1}}
\end{figure}

For comparison with Fig.~\ref{fig:PFs} and Fig.~\ref{fig:PF-Pf-T10} we also plot in 
Fig.~\ref{fig:PF-G-Pf-even0T10-r1} the power factor 
$\P_T$ computed from Eq.~(\ref{P_T}) for the Pfaffian state without bulk quasiparticles again at $T=T_0$, however this time 
for $r=1$. We see that again the sharp dips in the power factor correspond to 
the maxima of the conductance peaks but for $v_n=v_c$  the ratio of the two maxima of $\P_T$ around a 
conductance peak is obviously $1$. 
As mentioned before, the sharp zeros of the power factor can be used to determine experimentally 
the ratio $r=v_n/v_c$  because the CB peak pattern for all paired FQH states proposed for $\nu_H=5/2$ with 
even number of bulk quasiparticles  \cite{cappelli-viola-zemba,viola-stern} consists of a longer flux period
$\Delta\phi_2 =2+r$ and a shorter one $\Delta\phi_1 =2-r$, as shown in Fig.~\ref{fig:PF-G-Pf-even0T10-r1}, 
while that for the states with odd number of bulk quasiparticles  is equidistant, i.e.,  $\Delta\phi_1=\Delta\phi_2= 2$. 
This equidistant pattern of CB peaks could be used as a reference \cite{cappelli-viola-zemba,nayak-doppel-CB,thermal,viola-stern}. 
Since, according to Eq.~(\ref{shift}), the gate voltage $V_g$ is simply proportional to the AB flux $\phi$ we have that the ratio of the 
gate voltage periods is the same, i.e.,
$x=\Delta V_2/ \Delta V_1 =\Delta\phi_2 / \Delta\phi_1 \ge 1$ and therefore
\beq\label{r}
r= \lim_{T \to 0} 2\frac{(\Delta V_2/ \Delta V_1)-1}{(\Delta V_2/ \Delta V_1)+1}.
\eeq
For experimental purposes   the ratio $2(x-1)/(x+1)$ at temperatures $T\le T_0/2$ is very close to its zero-temperature value.

\begin{figure}[htb]
\centering
\includegraphics[bb=30 10 570 390,clip,width=10.5cm]{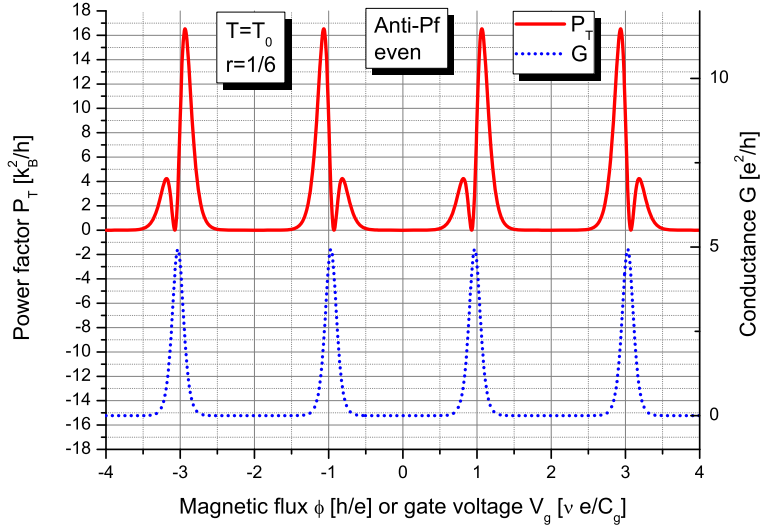}
\caption{The Power factor $\P_T$ and the conductance $G$ of a CB island for the Anti-Pfaffian state for $\nu_H=5/2$, 
without quasiparticles in the bulk,  with $r=1/6$, at temperature $T/T_0=1$. \label{fig:PF-G-APf-even-T10-r1-6}}
\end{figure}

We also plot in Fig.~\ref{fig:PF-G-APf-even-T10-r1-6} the power factor for the Anti-Pfaffian state with $r=1/6$ at $T=T_0$ 
computed from the partition function of Ref.~\cite{cappelli-viola,viola-stern}. We considered the partition function for the 
disorder-dominated 
phase of the Anti-Pfaffian state \cite{nayak-APf,rosenow-APf} with even number of bulk quasiparticles, in which the charged and 
neutral modes have already equilibrated \cite{kane-2-3,kane-fisher-equilib} and consequently the Hall conductance is universal,
as that of Eq.~(\ref{Z-ev}) with $\ch_0=[\chi^{(2)}_0]^{-1}$ and  $\ch_\omega=[\chi^{(2)}_2]^{-1}$, where $\chi_l^{(2)}$ are given 
in Eq.~(\ref{chi}). 
Again the sharp dips of $\P_T$ mark precisely the maxima of the conductance peaks
and can be used to determine precisely their positions.
\begin{figure}[htb]
\centering
\includegraphics[bb=40 20 560 390,clip,width=10.5cm]{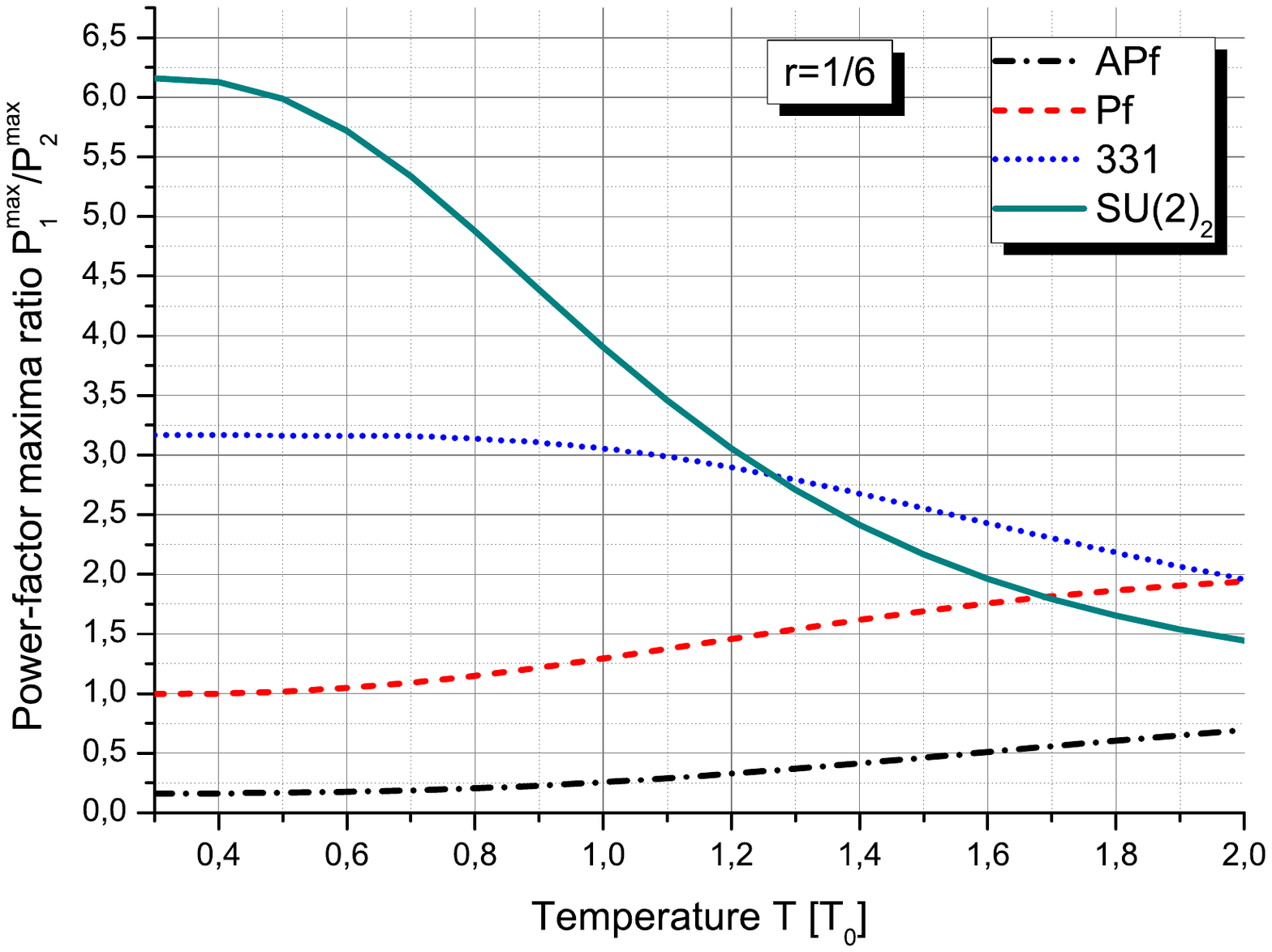}
\caption{Power-factor maxima ratio $P^{\max}_1/P^{\max}_2$ of a CB island for all paired states for $\nu_H=5/2$, without 
quasiparticles in the bulk, with $r=1/6$, as functions of  temperature $T/T_0$. \label{fig:PF-max-ratio-even-r1-6}}
\end{figure}
The plot in  Fig.~\ref{fig:PF-G-APf-even-T10-r1-6} has to be compared with the power factors of the other 
paired states given in Fig.~\ref{fig:PF-Pf-T10} and Fig.~\ref{fig:PFs}
The power factor for the Anti-Pfaffian state with even number of bulk quasiparticles 
is similar to that of the $SU(2)_2$ state though the places of higher and lower peaks, respectively the short and long 
periods in the gate voltage, are exchanged. This leads to different behavior  of the ratio $P_1^{\max}/P_2^{\max}$, as shown in 
Fig.~\ref{fig:PF-max-ratio-even-r1-6} which is also a clear signature of the Anti-Pfaffian or $SU(2)_2$ state.

Furthermore, the apparent asymmetries in $\P_T$ might allow to distinguish between the different states by measuring the ratio 
$P^{\max}_1/P^{\max}_2$  between the maxima  of $\P_T$  surrounding the first CB conductance peak with $\phi>0$. 
The plot of these ratios as functions of $T$, for the Pfaffian, 331, $SU(2)_2$ and the anti-Pfaffian states,  
are shown  in Fig.~\ref{fig:PF-max-ratio-even-r1-6}.
 Measuring $P^{\max}_1/P^{\max}_2$, as in \cite{gurman-2-3}, at three different temperatures, would be sufficient 
to determine experimentally one paired state among the others which is the best candidate to describe the universality class of the
 $\nu_H=5/2$ FQH state. Of course, the ratios of the maxima of the power factor around a conductance peak can be recalculated 
if the  ratio $r=v_n/v_c$, measured through Eq.~(\ref{r}), is  different from $1/6$.

\section{Discussion}
We demonstrated that the CFT partition functions of Coulomb blockaded FQH islands can be efficiently used to calculate 
the thermoelectric characteristics of the islands which could eventually distinguish between inequivalent FQH universality classes
with similar CB peaks patterns, at finite temperature even when $v_n/v_c<1$.

In this work we have considered only chiral FQH states. The only exception is the Anti-Pfaffian state for which we have 
used the partition function given in Refs.~\cite{cappelli-viola,viola-stern}. 
The reason is that for the chiral  FQH states all edge modes move in the 
same direction, which is determined by the direction of the magnetic field perpendicular to the FQH sample, and there is 
certainly a unitary rational CFT describing the edge states \cite{wen,fro-ker,ctz,read-CFT,cappelli-viola}.
For the non-chiral FQH states, such as the $\nu_H=2/3$ \cite{kane-2-3,gurman-2-3} and probably $\nu_H=5/2$  as well 
\cite{nayak-APf,rosenow-APf,neutral-5-2}, there might be counter-propagating 
neutral modes, or upstream modes, and it is not completely clear if conformal symmetry exists in the limit $v_n/v_c \to 1$,
so that the effective field theory partition function is unknown. 
If the $\nu_H=2/3$ FQH state is indeed the PH conjugate of the $\nu_H=1/3$ Laughlin state then it certainly has 
counter-propagating modes.
However, for most filling factors there are usually more than one candidate, and even if the experiments and numerical 
calculations favor one candidate there are phase transitions, etc; 
For example, in a recently proposed new Abelian candidate for $\nu_H=5/2$,
referred to as the 113 Halperin state \cite{113-5-2}, the  standard partition function is divergent because the $K$-matrix is 
not positive definite.
There are also some open problems, such as equilibration of counter-propagating modes in disorder-dominated phases, 
non-universality of the Hall conductance without equilibration, edge reconstruction, etc.
However, as soon as the partition function for any FQH state is fixed the method described here would allow to compute the 
thermopower, figure-of-merit,  power factor and conductances of a Coulomb-blockaded island inside this state.

It is worth mentioning that we consider the case $v_n=v_c$ as a ``zero approximation'' and admit that interactions could 
renormalize both velocities, so the  conformal symmetry exists exactly in this initial approximation. 
The important point here is that we expect that the structure of the Hilbert space 
of the edge states remain the same even after the renormalization so that we can use the same characters as partition function
and simply change the modular parameter $\tau$ in order to take into account the different velocities. There 
is a proof in Ref.~\cite{read-CFT} that when all modes propagate in the same direction there is certainly conformal symmetry on the edge.
The case with counter-propagating modes is not considered there and is unclear. If there is no CFT we don't know how to 
write the neutral partition functions, but if we knew them we could apply the approach described in this paper to compute the
thermoelectric  properties of the QD.

We conclude that the asymmetries in the power factor of Coulomb blockaded islands seem to be rather sensitive 
to the neutral degrees of freedom of the underlying edge states' effective conformal field theory and this could be used to 
determine experimentally which one of the candidate paired states describes best the fractional quantum Hall state 
at filling factor $\nu_H=5/2$.
\section*{Acknowledgments}
 I thank Andrea Cappelli, Guillermo Zemba and Ady Stern for helpful discussions.
This work has been partially supported by the Alexander von Humboldt Foundation under the Return Fellowship and 
Equipment Subsidies Programs and by the Bulgarian Science Fund under Contract No. DFNI-T02/6.
\appendix
\section{Additional details on Aharonov--Bohm twisting}
\label{app:AB}
Introducing AB flux $\phi=e(BA-B_0A_0)/h$ through the AB relation modifies the electron filed by 
$\psi_\el(z) \to z^{-\phi}\psi_\el(z)$, where $z=\e^{i\varphi}$ is the electron coordinate on the edge circle, see Sect.~2.8 in \cite{NPB-PF_k}.
This \textit{twisting} of the electron operator can be implemented by a conjugation \cite{NPB-PF_k} 
with a flux-changing operator $U_\beta$ (here $\beta\in \R$ is the twist parameter and should not be confused with the inverse temperature)
 defined by its commutation relations with the Laurent modes of the normalized \footnote{this means that the Laurent 
modes $J_n$ of the normalized charge density $J(z)=\sum_{n\in\Z} J_n z^{-n-1}$ satisfy $[J_n,J_m]=n\delta_{n+m,0}$ } charge density  
$J(z)=\sum_{n\in\Z} J_n z^{-n-1}$, namely $[J_n,U_{\beta}]=\beta U_\beta \delta_{n,0}$. It is not difficult to see that  $U_{\beta}$ 
acts on the electron as $\psi_\el(z) \to U_{\beta}\psi_\el(z) U_{-\beta}= z^{-\phi}\psi_\el(z)$ when the twist is 
$\beta=-\sqrt{\nu_H}\phi$ \cite{NPB-PF_k}. Then, the twisted 
electric charge is obtained by the same action  \cite{NPB-PF_k} $J^\el_0 \to U_\beta J^\el_0 U_{-\beta} =J^\el_0 + \nu_H \phi$ 
and  the twisted Hamiltonian, which is defined as   \cite{NPB-PF_k}
$H_{\mathrm{CFT}} \to U_\beta H_{\mathrm{CFT}} U_{-\beta} =H_{\mathrm{CFT}} + \Delta\varepsilon \phi J^\el_0 +  \Delta\varepsilon\nu_H \phi^2/2$,
reproduces Eq.~(\ref{H-N})  while the twisted partition function is expressed as in Eq.~(\ref{Z2}).

In order to make connection with the notation of Ref.~\cite{NPB-PF_k}, whose results for the AB transformation will be used below,
we denote $q=\e^{2\pi i\t} = \e^{-\beta\Delta\varepsilon}$ and  $\e^{2\pi i\z } = \e^{\beta\mu}$. 

The form of the twisted Hamiltonian $H_{\mathrm{CFT}}(\phi)$ defined in Eq.~(\ref{H-N})  
is typical (see, e.g. Eq.~(17) in Ref.~\cite{Viefers2004}) for two-dimensional interacting electron systems in which single-particle 
energies depend quadratically on the orbital momentum $n$ and hence depend quadratically on the AB flux after the shift 
\cite{Viefers2004}
$n \to n -\phi$.
Also we emphasize that while $N_\el$ denotes the electron number operator, in case when the  AB flux is non-zero, 
the operator  $N_{\imb}$, defined in (\ref{H-N}), is equal to the charge imbalance operator \cite{staring-CB}, 
i.e. it is the difference between the true electron number operator, which changes only by integers, and the externally induced 
charge $\nu_H \phi$, which varies continuously (either by external gate voltage or AB flux variation). To understand their relation
physically we note that  for $\phi=0$ the derivative of the Grand potential $\Omega=-k_B T \ln Z$ w.r.t. $\mu$ is
\beqa\label{dOmega}
\frac{\partial \Omega}{\partial \mu}=\frac{1}{Z}\sum_{s=0}^{n_H-1} \frac{\ch_{\omega^s * \Lambda}}{\eta(\t)}
\sum_{n\in \Z} \left(n_H n +s +\frac{l}{d_H}\right)  \times \nn
 \e^{-\beta\Delta\varepsilon \frac{n_Hd_H}{2}\left(n+\frac{l+sd_H}{n_Hd_H}\right)^2} \e^{2\pi i n_H\z\left(n+\frac{l+sd_H}{n_Hd_H}\right)},
\eeqa
and by definition should be equal to $-\la N_\imb\ra_{\beta,\mu}$. 
When we introduce AB flux by the shift (\ref{shift}) the $K$ functions  in $Z^{l,\Lambda}$
transform as 
\[
K_{l+sd_H}(\t,n_H\z; n_Hd_H) \to K_{l+sd_H}(\t,n_H(\z+\phi\t); n_Hd_H) .
\]
However, the latter is equal to
 $K_{l+sd_H+n_H\phi}(\t,n_H\z; n_Hd_H)$ due to the $K$-function identity (\ref{K-shift})
 i.e., the effect of adding AB flux is simply to shift the $K$-function index $l \to l+n_H \phi$. Then
 shifting $l$ in (\ref{dOmega}) yields in the sum an extra term proportional to $\phi$ which after the average gives rise to
\beq \label{dOmega2}
\frac{\partial \Omega_\phi}{\partial \mu} = -\la N_\el(\phi)\ra_{\beta,\mu} +\nu_H\phi =-\la N_\imb(\phi)\ra_{\beta,\mu},
\eeq
that explains Eqs.~(\ref{N})and (\ref{H-N}). It is interesting to note that the operator $Q_\imb\equiv -N_\imb$ 
coincides with the zero mode of the twisted $u(1)$ current $\pi_\phi (J_0^{\el})$ defined in \cite{NPB-PF_k},
which is precisely the operator that appears in the twisted partition function (\ref{Z2}) as coupled to $\mu$.
The electron number $N_\el$ on the edge of a Pfaffian CB island without quasiparticles in the bulk, computed numerically from 
Eq.~(\ref{N}) in the text,  is shown in Fig.~\ref{fig:G-N} together with  the electric conductance $G$ of the island
computed from Eq.~(\ref{G})  for $l=0$ and $\Lambda=0$, at $\mu=0$ and 
$r=1$, as functions of the magnetic flux.

\section{Fixing chemical potentials $\mu_N$ and $\mu_{N+1}$}
\label{app:mu}
Now we continue with the explanation why we choose $ \mu_N=-\Delta \varepsilon /2$ and  $\mu_{N+1}=\Delta \varepsilon /2$.
First, let us see why the partition function (\ref{Z_l-L-phi}) is independent of the bulk chemical potential $\mu_0$ but depends 
only on the edge part $\mu=\mu_{\mathrm{tot}}-\mu_0$ as stated in the text. 
The electron number average derived in general from (\ref{dOmega2}) with a total chemical potential $\mu_{\mathrm{tot}}=\mu_0+\mu$
\[
 \la N(\phi)\ra_{\beta,\mu} = -\frac{\partial \Omega_{\phi}}{\partial \mu} +\nu_H\left( \frac{\mu_{\mathrm{tot}}}{\Delta\varepsilon} +\phi\right)
=N_0+ \la N_\el(\phi)\ra_{\beta,\mu},
\]
contains a  bulk term $\nu_H \mu_0/\Delta\varepsilon$ and an edge term $\nu_H\mu/\Delta\varepsilon -\partial\Omega_\phi/\partial \mu$.
Because the bulk term, corresponding to $\phi=0$, must be equal to $N_0$ we find 
$\mu_0=d_H\Delta\varepsilon N_0/n_H$. 
Next we assume,  as in the text, that the number of electrons in the bulk is $N_0=n_H n_0$,
where $n_0$ is a positive integer, so that the bulk chemical potential becomes $\mu_0/\Delta\varepsilon  = n_0 d_H $.
Now, if we substitute $\mu_{\mathrm{tot}}$ into the partition function (\ref{Z_l-L-phi})  we see that the latter is independent of the bulk
chemical potential $\mu_0$ because the sum over $n$ is invariant with respect to the shift $n\to n+n_0$, i.e.,
$Z_{\phi}^{l,\Lambda}(\t,{\mu_0+\mu})=Z_{\phi}^{l,\Lambda}(\t,{\mu})$. Therefore the edge partition function (\ref{Z_l-L-phi}), 
as well as all thermodynamic averages, depend only on 
the edge part $\mu$ of the total chemical potential $\mu_{\mathrm{tot}}=\mu_0+\mu$.

In order to determine the values of $\mu_N$ and $\mu_{N+1}$, which are needed for the computation of the thermodynamic 
averages of the energy $H_{\mathrm{CFT}}(\phi)$ and the electron number $N_\el(\phi)$, we argue that the difference 
$\mu_{N+1}-\mu_{N}$ corresponds to the difference between the energy of the last occupied single-particle state in the 
CB island and the first available unoccupied single-particle state. This difference is proportional to the flux difference between 
the two states, i.e. $\mu_{N+1}-\mu_{N}=\phi\Delta\varepsilon $. The first unoccupied single-particle state in the QD
can be obtained from the last occupied one by the Laughlin spectral flow \cite{cz}: we apply the Laughlin argument \cite{laughlin}
to the last occupied single particle state by changing adiabatically the AB flux threading the electron disk from $0$ to $1$; when the 
flux becomes $1$ the discrete spectrum of the Hamiltonian $H_{\mathrm{CFT}}(\phi)$ becomes the same as the spectrum for $\phi=0$
while the single-electron states are mapped onto themselves, i.e., if at $\phi=0$ an electron is described by the (unnormalized) wave 
function $z^l \e^{-|z|^2/4}$,  then at $\phi=1$
it would have the wave function  $z^{l+1} \e^{-|z|^2/4}$. Therefore the difference between the two chemical 
potentials corresponding to 
the last occupied and the first unoccupied single-particle states is exactly $\mu_{N+1}-\mu_{N}=\Delta\varepsilon$. 
Put another way, the Laughlin spectral flow, which transforms the last occupied single-particle level to the first unoccupied one,
is expressed by the (modular) transformation \cite{cz} $\z \to \z+\tau$ and it can be implemented by Eq. (\ref{shift}) with $\phi=1$. 
Taking into account that
$\z = (\mu/\Delta\varepsilon)\t$, as defined in the text before Eq.~(\ref{shift}), we conclude that the Laughlin spectral flow is equivalent to 
$(\mu/\Delta\varepsilon)\to (\mu/\Delta\varepsilon)+1$, or $\mu \to \mu +\Delta\varepsilon$ so that $\Delta\mu =\Delta\varepsilon$.
Next, assuming that $\mu_{N}+\mu_{N+1}=0$, which is equivalent to fixing the QD in the center of a CB valley for $\phi=0$,
we finally obtain  $\mu_N=-\Delta\varepsilon/2$ and $\mu_{N+1}=\Delta\varepsilon/2$ as in Eq.~(\ref{mu_NN1}). 
It is worth stressing that $\mu_{N+1}-\mu_{N}$
is independent of the neutral degrees of freedom of the electrons in the QD, hence it is independent of the ratio $r=v_n/v_c$ of the 
Fermi velocities of the charged and neutral edge modes. It is also independent of the filling factor $\nu_H$.
Note however that, except for $\phi=0$, $\mu_N$ is not the true chemical potential, because it is coupled in the partition function (\ref{Z2})
 to the charge imbalance operator $N_\imb$ defined in (\ref{H-N}) instead of the particle number $N_\el$.
That is why, $\mu_{N+1}-\mu_{N}$ is not equal to the addition energy 
$\Delta\mu^{\mathrm{phys}}=\mu^{\mathrm{phys}}_{N+1}-\mu^{\mathrm{phys}}_{N}$, 
where $\mu^{\mathrm{phys}}_{N}$ is the true (physical) 
chemical potential. On the other hand, the addition energy $\Delta\mu^{\mathrm{phys}}_{N}$ corresponds to the energy
spacing between CB peaks and it can be interpreted as the difference between the energies of the ground states with
$N+1$ and $N$ electrons and certainly depends on the neutral degrees of freedom and on the filling factor $\nu_H$.
\section{Electric charge and Luttinger liquid number for general FQH states}
 The modular parameter $\z$ in Eq.~(\ref{Z_l-L-phi}) carries an additional multiplicative 
factor $n_H$, the numerator of the filling factor $\nu_H=n_H/d_H$, which can be understood as follows \cite{LT9}:  
the $K$ functions (\ref{K}) entering the full partition functions (\ref{Z_l-L-phi}) are actually the partition functions for the Luttinger liquid 
which can be described by a chiral boson with a compactification radius \cite{cz} $R_c=1/m$, where  
 $m=n_H d_H$. This is because the $\uu$ component of the electron field $\e^{i\phi(z)/\sqrt{\nu_H}}$, 
which is fixed by the requirement 
to have electric charge $1$, has a statistical angle $\theta/2\pi=d_H/n_H$ that is not integer for $n_H>1$. 
Therefore, for $n_H>1$ we need to consider
 a smaller chiral subalgebra containing only clusters of $n_H$ electrons \cite{LT9}, which can be generated by 
$\e^{i n_H\phi(z)/\sqrt{\nu_H}}\simeq \e^{i \sqrt{n_H d_H}\phi(z)}$.
Theses fields have integer statistics and are local,  however their corresponding $\uu$ compactification radius 
$n_H d_H$ is bigger \cite{LT9}. 
The electric charge operator $Q$ can be expressed in terms of the normalized chiral $\uu$ charge $J_0$
 by $Q=\sqrt{\nu_H} J_0$. On the other hand the normalized charge $J_0$ is related to the Luttinger liquid number operator 
$N=-J_0/\sqrt{m}$, whose spectrum appears in the Luttinger liquid partition function (\ref{K}) as a conjugate of $\z$,
so that combining both we can express $Q$ by $N$ as follows
\[
Q=\sqrt{\frac{n_H}{d_H}} J_0 = -\sqrt{\frac{n_H}{d_H}} \sqrt{n_H d_H} N = -n_H N.
\]
This implies that the modular parameter $\z$ in the partition function (\ref{Z_l-L-phi}) must be multiplied by $n_H$.
This detail has one important consequence: the charge index $l+sd_H$, which is divided by $n_Hd_H$ in (\ref{Z_l-L-phi}), is deformed by 
adding $n_H \phi$ in presence of extra AB flux. Therefore the AB flux enters (\ref{dOmega2}) as $\phi n_H/d_H$ which
is necessary  for the correct implementation of the charge-flux relation in general FQH states \cite{fro-stu-thi,fro2000,LT9}.
\section*{References}
\bibliography{my,TQC,FQHE,Z_k,CB}
\end{document}